**The IBEX Knowledge-Base: Achieving more together with open science**

Andrea J. Radtke[1,*], Ifeanyichukwu Anidi[2], Leanne Arakkal[3], Armando Arroyo-Mejias[3], Rebecca T. Beuschel[3,36], Katy Börner[4], Colin J. Chu[5], Beatrice Clark[3], Menna R. Clatworthy[6], Jake Colautti[7], Joshua Croteau[8], Saven Denha[7], Rose Dever[9], Walderez O. Dutra[10], Sonja Fritzsche[11], Spencer Fullam[12], Michael Y. Gerner[13], Anita Gola[14], Kenneth J. Gollob[15], Jonathan M. Hernandez[16], Jyh Liang Hor[3], Hiroshi Ichise[3], Zhixin Jing[3], Danny Jonigk[17,18], Evelyn Kandov[3,37], Wolfgang Kastenmüller[19], Joshua F.E. Koenig[7], Aanandita Kothurkar[5], Alexandra Y. Kreins[20,21], Ian Lamborn[3], Yuri Lin[16], Katia Luciano Pereira Morais[15], Aleksandra Lunich[2], Jean C. S. Luz[22], Ryan B. MacDonald[5], Chen Makranz[23], Vivien I. Maltez[24], Ryan V. Moriaty[25], Juan M. Ocampo-Godinez[20,26,27], Vitoria M. Olyntho[7], Kartika Padhan[1], Kirsten Remmert[16], Nathan Richoz[6], Edward C. Schrom[3], Wanjing Shang[3], Lihong Shi[28], Rochelle M. Shih[3], Emily Speranza[29], Salome Stierli[30], Sarah A. Teichmann[31,32], Tibor Z. Veres[3], Megan Vierhout[7], Brianna T. Wachter[33], Adam K. Wade-Vallance[3], Margaret Williams[2], Nathan Zangger[34], Ronald N. Germain[1,3,*], and Ziv Yaniv[35,*]

Corresponding authors: Andrea J. Radtke (andrea.radtke@nih.gov), Ziv Yaniv (zivyaniv@nih.gov), and Ronald N. Germain (rgermain@niaid.nih.gov)

[1]Lymphocyte Biology Section and Center for Advanced Tissue Imaging, Laboratory of Immune System Biology, National Institute of Allergy and Infectious Diseases, National Institutes of Health, Bethesda, MD, USA

[2]Critical Care Medicine and Pulmonary Branch, National Heart, Lung and Blood Institute, National Institutes of Health, Bethesda, MD, USA

[3]Lymphocyte Biology Section, Laboratory of Immune System Biology, National Institute of Allergy and Infectious Diseases, National Institutes of Health, Bethesda, MD, USA

[4]Department of Intelligent Systems Engineering, Indiana University, Bloomington, IN, USA

[5]UCL Institute of Ophthalmology and NIHR Moorfields Biomedical Research Centre, London, UK

[6]Cambridge Institute for Therapeutic Immunology and Infectious Diseases, University of Cambridge Department of Medicine, Molecular Immunity Unit, Laboratory of Molecular Biology, Cambridge, UK

[7]McMaster Immunology Research Centre, Schroeder Allergy and Immunology Research Institute, Department of Medicine, Faculty of Health Sciences, McMaster University, Hamilton, ON, Canada

[8]Department of Business Development, BioLegend Inc., San Diego, CA, USA

[9]Functional Immunogenomics Unit, National Institute of Arthritis and Musculoskeletal and Skin Diseases, National Institutes of Health, Bethesda, MD, USA

[10]Laboratory of Cell-Cell Interactions, Department of Morphology, Institute of Biological Sciences, Universidade Federal de Minas Gerais, Belo Horizonte, MG, Brazil

[11]Max-Delbrueck-Center for Molecular Medicine in the Helmholtz Association (MDC), Spatial Proteomics Group, Berlin, Germany

[12]Division of Rheumatology, Rush University Medical Center, Chicago, IL, USA

[13]Department of Immunology, University of Washington School of Medicine, Seattle, WA, USA

[14]Robin Chemers Neustein Laboratory of Mammalian Cell Biology and Development, The Rockefeller University, New York, NY, USA

[15]Center for Research in Immuno-oncology (CRIO), Hospital Israelita Albert Einstein, Sao Paulo, SP, Brazil

[16]Surgical Oncology Program, National Cancer Institute, National Institutes of Health, Bethesda, MD, USA

[17]Institute of Pathology, Aachen Medical University, RWTH Aachen, Aachen, Germany

[18]German Center for Lung Research (DZL), Biomedical Research in Endstage and Obstructive Lung Disease Hannover (BREATH), Hannover, Germany

[19]Würzburg Institute of Systems Immunology, Max Planck Research Group at the Julius-Maximilians-Universität Würzburg; Würzburg, Germany

[20]Infection Immunity and Inflammation Research and Teaching Department, University College London Great Ormond Street Institute of Child Health, London, UK

[21]Great Ormond Street Hospital for Children NHS Foundation Trust, Department of Immunology and Gene Therapy, London, UK

[22]Viral Vector Laboratory, Cancer Institute of São Paulo, University of São Paulo, SP, Brazil

[23]Neuro-Oncology Branch, National Cancer Institute, National Institutes of Health, Bethesda, MD, USA

[24]Division of Allergy, Immunology and Rheumatology, Department of Pediatrics, University of California San Diego, La Jolla, CA, USA

[25]Department of Cellular and Developmental Biology, Northwestern University, Chicago, IL, USA

[26]Laboratorio de Bioingeniería de Tejidos, Departamento de Estudios de Posgrado e Investigación, Universidad Nacional Autónoma de México, Mexico City, Mexico

[27]Laboratorio de Inmunoquímica I, Departamento de Inmunología, Escuela Nacional de Ciencias Biológicas, Instituto Politécnico Nacional, Mexico City, Mexico

[28]Laboratory of Immune System Biology, National Institute of Allergy and Infectious Diseases, National Institutes of Health, Bethesda, MD, USA

[29]Florida Research and Innovation Center, Cleveland Clinic Lerner Research Institute, Port Saint Lucie, FL, USA

[30]Institute of Anatomy, University of Zurich, Zurich, Switzerland

[31]Cambridge Stem Cell Institute, Jeffrey Cheah Biomedical Centre, Puddicombe Way, Cambridge Biomedical Campus, Cambridge, UK

[32]Department of Medicine, University of Cambridge, Cambridge, UK

[33]Laboratory of Clinical Immunology and Microbiology, National Institute of Allergy and Infectious Diseases, National Institutes of Health, Bethesda, MD, USA

[34]Institute of Microbiology, ETH Zurich, Zurich, Switzerland

[35]Bioinformatics and Computational Bioscience Branch, National Institute of Allergy and Infectious Diseases, National Institutes of Health, Bethesda, MD, USA

[36]Current address for Rebecca T. Beuschel: Division of Rheumatology, Inflammation, and Immunity, Department of Medicine, Brigham and Women's Hospital and Harvard Medical School, Boston, MA, USA

[37]Current address for Evelyn Kandov: Department of Pharmacology, Vanderbilt Brain Institute, Vanderbilt Center for Addiction Research, Vanderbilt University, Nashville, TN, USA

**Abstract**

Iterative Bleaching Extends multipleXity (IBEX) is a versatile method for highly multiplexed imaging of diverse tissues. Based on open science principles, we created the IBEX Knowledge-Base, a resource for reagents, protocols and more, to empower innovation.

**Manuscript**

*The power and pitfalls of multiplexed tissue imaging*

Multiplexed tissue imaging is a powerful approach for studying the spatial organization and cellular composition of intact tissues at single cell resolution. The last decade has seen a rapid expansion in the development and commercialization of advanced spatial biology techniques. These methods include technologies that probe RNA molecules using imaging-based approaches or spatial barcoding techniques. In addition, proteins may be targeted with antibodies applied to thin sections as well as thick tissue volumes using a variety of multiplexed antibody-based imaging approaches[1]. These methods vary in the optical resolution, tissue volume, and number and type of targets (RNA, protein, or both) that can be imaged in a single tissue preparation[2]. As with any rapidly evolving field, the technical specifications of a given method are constantly improving, enhancing the value of these approaches. Multiplexed tissue imaging has been especially informative for visualizing and quantifying cell-cell interactions, identifying rare cells, evaluating spatial relationships among cells, and providing new insights into higher level tissue organization. These technologies have been foundational for the construction of single cell atlases and the study of naturally occurring cancers using samples from clinical trials and experimental models of disease. Despite their considerable promise, several challenges prevent the widespread adoption of spatial biology technologies. First and foremost, the majority of these methods require expensive equipment and consumables that may not be available in all research settings. Extensive expertise is also needed to optimize tissue collection, validate reagents, acquire images, and analyze data[1].

*IBEX: An open and versatile method for highly multiplexed imaging*

To provide a robust and widely usable solution for highly multiplexed imaging, we developed the Iterative Bleaching Extends multipleXity (IBEX) method[3, 4]. This method achieves high parameter imaging (>65 markers) in a single tissue section (5-30 μm) using cyclic rounds of antibody labeling and dye inactivation. Following image acquisition, individual images are registered, pixel-to-pixel, into one composite image using open-source software[5]. Since December 2020[3], IBEX has been adopted by dozens of international scientists from fields as diverse as immunology, developmental biology, comparative anatomy, and cancer biology. Furthermore, IBEX has been used to evaluate tissues obtained from humans, mice, non-human primates, canines, and zebrafish. More importantly, these advances reflect the community's ability to both adopt and expand the original IBEX method to different applications and laboratory settings. As a result, we have collectively overcome common challenges, developed workflows for automated imaging and immunolabeling, and incorporated new reagents to acquire high quality imaging datasets for a variety of studies.

*Motivation and design for IBEX Knowledge-Base*

From the beginning, we have strived to share knowledge related to each stage of the multiplexed imaging workflow: sample preparation, antibody selection, antibody validation, panel design, image alignment, image processing, data analysis, and publication of results via open data repositories and scholarly publications. This effort was born out of a desire to reduce the significant time, resources, and expertise required to implement IBEX and other multiplexed imaging techniques[1, 6]. To achieve this aim, we established the IBEX Knowledge-Base, a central resource for reagents, protocols, data, software, and information related to IBEX and other spatial biology methods including non-iterative, standard tissue imaging (Multiplexed 2D Imaging), IBEX imaging with Opal dyes (Opal-plex), thick volume imaging achieved through clearing enhanced 3D (Ce3D)[7], and integration of Ce3D and IBEX (Ce3D-IBEX) to obtain highly multiplexed imaging of thick samples (>300 µm). We anticipate the number of methods supported by the IBEX Imaging Community to grow and include unique extensions of the protocol for the detection of novel chemistries and nucleic acid probes.

The IBEX Knowledge-Base is designed around three facets common to FAIR data and open-source software development: a source/data repository, a static website, and an archive for source data (Yaniv *et al.*, in preparation). The first facet, the IBEX Knowledge-Base GitHub repository, stores source data and scripts used to generate the static website (https://github.com/IBEXImagingCommunity/ibex_imaging_knowledge_base). Furthermore, the GitHub ecosystem provides support for automatic data validation, website creation and hosting, issue reporting, as well as a discussion forum. These latter two utilities provide an open, transparent venue for discussing issues and questions related to the IBEX Knowledge-Base and multiplexed tissue imaging, respectively. There are several benefits to this approach such as increased trust in the data, reduced time in answering the same or similar questions in closed communication channels, and access to a wider pool of expertise. The second facet, the static website (https://ibeximagingcommunity.github.io/ibex_imaging_knowledge_base), is automatically generated with every update to the IBEX Knowledge-Base via a GitHub pull request. The IBEX Imaging Community website was designed to provide a user-friendly platform for browsing the current stage of knowledge and, unlike scientific publications, is constantly evolving with each contribution. The third and final facet of the IBEX Knowledge-Base is publication of an authoritative, citable, archival version through the generalist repository Zenodo[8]. By publishing through Zenodo, the IBEX Knowledge-Base is assigned a persistent digital object identifier, providing a mechanism for members to be rewarded with authorship for their contributions. In contrast to the static website, updates to the Zenodo entry are not continuous but are issued as versioned releases following a significant update.

*Guiding principles*

The IBEX Knowledge-Base was founded on five guiding principles (Figure 1). First, we are better together and, importantly, achieve more together by adopting a mindset of *shared ownership*. For this reason, everyone who contributes knowledge in the form of a reagent resource, antibody validation image, protocol created expressly for the community, or video tutorial is named as an author on the Zenodo dataset and static website. Our second principle is *failure teaches success*. Unlike publications in which only successful work is described, the goal of the IBEX Knowledge-Base is to document both successful and failed work. By sharing failures, we advance science at a faster pace, reduce financial costs, and instill confidence in the resulting data. While it's good to learn from our own failures, it's better to learn from our failures *and* the failures of others. Accordingly, our discussion forum provides an opportunity for community members to learn from others. Our third principle, *stewardship and democratization*, is rooted in the open science principles of data sharing, equity, and inclusion. Beyond sharing recommended reagents, we actively encourage the communication of unsatisfactory reagents to prevent other researchers from wasting time and resources. Through stewardship we make science more equitable, reduce the significant cost of validating antibodies[1, 6], and empower scientists from around the world to perform multiplexed imaging. Fourth, members of the IBEX Knowledge-Base are distinguished by a *commitment to excellence*. To achieve this goal, we collect metadata critical for the performance of a reagent including the target species, tissue preservation method, antigen retrieval conditions, detergent used in blocking and immunolabeling buffers, and information on the best conjugate or antibody clone. We also designed a mechanism for self-correction whereby members of the community can "agree" or "disagree" with reagent entries using their Open Researcher and Contributor ID (ORCID). Like *PLos Biology*[9], we believe in the importance of being second because reproducible science is good science. Our most current state of knowledge, found on the static website, reports sixteen reagent entries replicated by two independent authors and one reagent entry replicated by three experts. More than a year after its launch, we celebrated our first disagreement regarding an antibody that labels anticipated cell types as well as unusual cell types in the mouse lymph node. We welcome you to join the conversation in the post titled, "Our first disagreement" in the discussion forum. This contribution exemplifies our fifth and final principle by demonstrating the *power of iteration*, particularly as it applies to correcting and refining our collective state of knowledge. With each addition to the IBEX Knowledge-Base, our knowledge about sample preparation, reagents, and many other aspects of multiplexed imaging and analysis grows (Figure 1).

*An open invitation to use and contribute*

The evolving state of knowledge, reflected by the plus signs associated with the number of sample preparations, protocols, videos, reagents, etc. (Figure 1), is made possible by contributors like you! To date, more than 25 contributors from Brazil, Canada, Germany, Mexico, Switzerland, the United Kingdom, and the United States have shared their expertise with the community. The IBEX Knowledge-Base operates under the Creative Commons Attribution 4.0 license which allows anyone to use the resources collected here with attribution. Before embarking on multiplexed tissue imaging, we invite you to use the Knowledge-Base to identify the best way to prepare your samples based on protocols, videos, publications, and support offered via the discussion forum. There are several ways to use the "Reagent Resources" tab on the IBEX Imaging Community website to find suitable reagents for your study. The most common approach is to use the filter function to find community-validated antibodies that are "recommended Yes" for your target species, tissue preservation method (formalin-fixed paraffin embedded (FFPE), fixed frozen, etc.), and antigen retrieval conditions. Another option is to use the "Reagent Resources" tab and community provided publications to identify what other members are examining in the same or similar tissues. Lastly, the extensive list of vendors (35+) may help investigators find where to purchase antibodies for non-traditional experimental animal model systems, e.g., Zebrafish International Resource Center (ZIRC). Finally, in accordance with our guiding principles we encourage you to return to the IBEX Knowledge-Base and add your failures and successes, celebrate your accomplishments, and share your knowledge with others. Your success is our success!

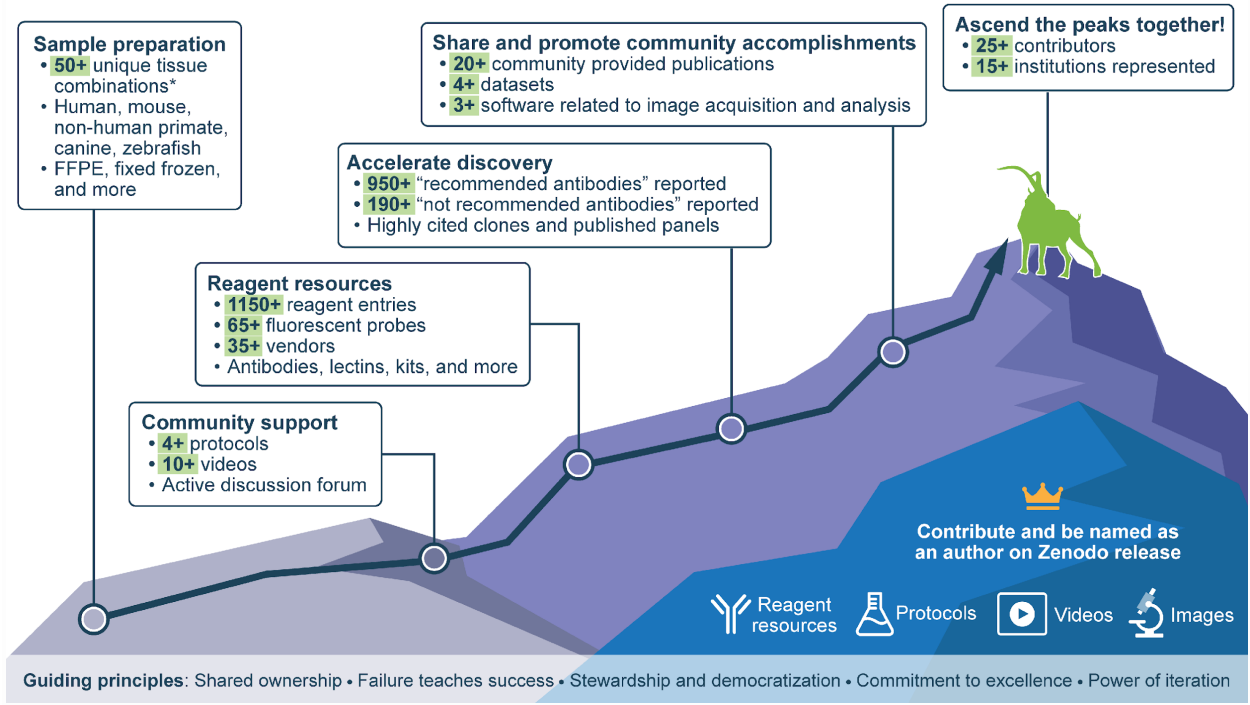

**Figure 1. The IBEX Knowledge-Base is a central portal for IBEX and related multiplexed tissue imaging techniques.**

The IBEX Knowledge-Base is an open, global repository providing information related to IBEX and other spatial biology methods. The evolving state of knowledge is reflected by the plus signs associated with the information found here. The 50+ unique tissue combinations are calculated using details related to the target species, tissue preservation method, target tissue, and tissue state, e.g., infected with a particular pathogen. The crown denotes contributions leading to authorship on the Zenodo release (some restrictions apply). The ibex (goat) climbing the mountain is symbolic of the method's namesake.

### Acknowledgments

We are deeply appreciative of Arlene Radtke for her encouragement, proof-reading, and assistance with antibody metadata fields. This work was supported by the Division of Intramural Research, NIAID, NIH, Center for Cancer Research, NCI, NIH, NHLBI, NIH, and NIAMS, NIH under the following grants: 1ZIAAI000758-26 and 1ZIAAI000545-35. KB is supported by the NIH Common Fund through the Office of Strategic Coordination/Office of the NIH Director under award OT2OD033756, the SenNet CODCC under award number U24CA268108, NIDDK under award U24DK135157, the KPMP grant U2CDK114886, and the CIFAR MacMillan Multiscale Human program. CJC is supported as a Wellcome Trust Clinical Research Career Development Fellow (224586/Z/21/Z). MRC and the Clatworthy lab (NR) are supported by a Wellcome Investigator Award (220268/Z/20/Z), the National Institute of Health Research (NIHR) Cambridge Biomedical Research Centre (NIHR203312), and the NIHR Blood and Transplant Research Unit in Organ Donation and Transplantation (NIHR203332), a partnership between NHS Blood and Transplant, the University of Cambridge and Newcastle University. The views expressed are those of the authors and not necessarily those of the NIHR or the Department of Health and Social Care. JC, SD, and JFEK are supported by a peer-reviewed Food Allergy Research Grant jointly funded by the CIHR Institute of Infection and Immunity (CIHR-III), CIHR Institute of Circulatory and Respiratory Health (CIHR-ICRH), and the Canadian Allergy, Asthma and Immunology Foundation (CAAIF), and by the Walter and Maria Schroeder Foundation. MYG is supported by NIH grant R01AI134713 and NIH contract 75N93019C00070. AG is funded by Damon Runyon Cancer Research Foundation (National Mah Jongg League Fellowship (DRG 2409-20)). KJG and KLPM are supported by CNPq, FAPEMIG, FAPESP (#2021/00408-6), Instituto Nacional de Ciencia e Tecnologia em Doenças Tropicais (INCT-DT). DJ is supported by the grant of the European Research Council (ERC); European Consolidator Grant, XHale (Reference #771883). WK is supported by the European Research Council (ERC) (819329-STEP2). AK and RBM are supported by a Biotechnology and Biological Sciences Research Council (BBSRC) David Phillips fellowship (BB/S010386/1). AYK is supported by the Wellcome Trust (222096/Z/20/Z). All research at GOSH is supported by the National Institute of Health Research (NIHR) GOSH Biomedical Research Centre (BRC). JCSL is supported by the Sao Paulo Research Foundation (FAPESP fellowship 2023/01697-7). VIM is supported by the NIGMS MOSAIC K99/R00 4R00GM147841-02. JMOG is supported by the National Polytechnic Institute (IPN) of Mexico (ID: DRI/DII/0445/2024) and the National Council of Humanities, Science and Technology (Conahcyt) (ID: B210344 and CVU: 10008) with fundings for research at UCL-Great Ormond Street Institute Child Health (UCL-GOSH). MV was supported by the Canadian Institutes of Health Research (CIHR) Doctoral Award (Grant No. 170793) and the Ontario Graduate Scholarship (OGS) Program. Z.Y. is supported by the BCBB Support Services Contract HHSN316201300006W/75N93022F00001 to Guidehouse Inc.

**Competing interests**

SAT is a remunerated Scientific Advisory Board member of Qiagen, Foresite Labs, OMass Therapeutics, and a consultant and equity holder of TransitionBio and EnsoCell, and a non-executive board director of 10x Genomics, as well as part-time employee of GlaxoSmithKline. JC is an employee and stakeholder of BioLegend (revvity inc.). All other authors declare no competing interests.